\documentclass[aip,apl,preprint]{revtex4-1}

\usepackage{graphicx}
\usepackage{dcolumn}
\usepackage{bm}
\usepackage{amsmath}
\usepackage{amssymb}
\usepackage{color}

\begin{document}

\title{The limits of flexoelectricity in liquid crystals}

\author{F.~Castles}
\author{S.~M.~Morris}
\author{H.~J.~Coles}
\affiliation{Centre of Molecular Materials for Photonics and Electronics, Department of Engineering, University of Cambridge, 9 JJ Thomson Avenue, Cambridge CB3 0FA, United Kingdom}


\begin{abstract}
The flexoelectric conversion of mechanical to electrical energy in nematic liquid crystals is investigated using continuum theory.  Since the electrical energy produced cannot exceed the mechanical energy supplied, and vice-versa, upper bounds are imposed on the magnitudes of the flexoelectric coefficients in terms of the elastic and dielectric coefficients.  For conventional values of the elastic and dielectric coefficients, it is shown that the flexoelectric coefficients may not be larger than a few tens of pC/m.  This has important consequences for the future use of such flexoelectric materials in devices and the related energetics of distorted equilibrium structures.
\end{abstract}


\maketitle

Materials composed of small elongated organic molecules may form a nematic liquid crystal (LC) phase, with molecules aligned, on average, along a common direction $\pm{\bf n}$.\cite{gennesbook}  In the ground state, ${\bf n}$ is uniform, and an elastic energy cost is associated with distortions of ${\bf n}$.  Flexoelectricity is a coupling between certain such distortions and electric polarization; an applied electric field may induce distortion, and distortion may induce polarization.\cite{meyer}  The effect is of interest in devices such as display panels, \cite{patel,zbdpatent,patent,broughton,coles,davidson,castlespre,chen2,carbone,salter2,castlesjsid} and electro-mechanical transducers for sensing or energy-harvesting applications.\cite{jaklipatent}  More fundamentally, flexoelectricity has recently been studied as a factor influencing the energetics of distorted equilibrium structures.\cite{castlesprl,porenta}  The strength of the coupling---described by the flexoelectric coefficients---is of primary importance in such considerations.

Considerable research has been devoted to developing highly flexoelectric ``bimesogenic'' materials (flexoelectric coefficients of magnitude of order $\sim 10$~pC/m).\cite{coles,morris}  These reduce the driving voltage for certain electro-optic devices,\cite{coles} and produce unusually stable blue phases.\cite{nature,castlesprl}  It is important to ask how far the development can continue; what are the maximum flexoelectric coefficients practically achievable?  Helfrich\cite{helfrich} has shown theoretically---using phenomenological and molecular-statistical approaches---that the magnitude of the flexoelectric coefficients should be $\leq\sim10$~pC/m, indicating the maximum has almost been reached.  However, independently of the above investigations, Harden \textit{et al}.\cite{harden,harden2} reported the measurement of ``giant'' flexoelectricity: coefficients 3--4 orders of magnitude greater than those known previously ($\sim 10\,000$~pC/m), and in apparent conflict with Helfrich's theory.  Some discussion has developed as to how this significant difference between theory and experiment may be reconciled.\cite{harden,kumar}  From conservation of energy considerations, we will show that there is a clear fundamental limit to the flexoelectric coupling strength; the electrical energy induced cannot exceed the mechanical energy supplied to distort ${\bf n}$, and vice-versa.  By considering theoretically the energy conversion for the principal director distortions, upper bounds on the magnitudes of the flexoelectric coefficients are derived.

The elastic energy cost of distortions of ${\bf n}$ may be accurately described using the Oseen-Frank continuum theory.\cite{gennesbook,oseen,frank}  Ignoring surface terms (which do not appear for the distortions considered here), the free energy per unit volume with respect to that of the undeformed state is\cite{oseen,frank,gennesbook}
\begin{equation} \label{eq:fenergy}
F_\textrm{d} = \frac{1}{2}K_{1}S^2 + \frac{1}{2}K_2 T^2 + \frac{1}{2}K_3 B^2,
\end{equation}
where ${\bf S}={\bf n}\left(\bm\nabla \cdot {\bf n}\right)$, $T={\bf n} \cdot
\bm\nabla \times {\bf n}$, and ${\bf B}=(\bm\nabla \times \mathbf{n}) \times{\bf n}$, are splay, twist, and bend distortions, respectively.  ($S^2\equiv {\bf S}\cdot{\bf S}$ and $B^2\equiv {\bf B}\cdot{\bf B}$.)  $K_1$, $K_2$, and $K_3$ are the associated elastic coefficients, which are positive in the conventional nematic phase considered here.\cite{gennesbook}  Splay and bend distortions induce a flexoelectric polarization
\begin{equation}
{\bf P}_\textrm{d}=e_1{\bf S}+e_3{\bf B},
\end{equation}
(but twist does not).\cite{meyer}  This defines $e_1$ and $e_3$, the splay and bend flexoelectric coefficients respectively (using the notation of Ref.~\onlinecite{gennesbook}).

\begin{figure}
\includegraphics[width=0.42\textwidth]{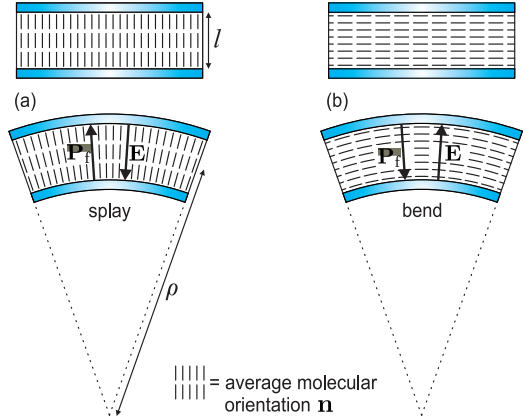}
\caption{\label{fig:limits} (Color online) Distortions of (a) pure splay, and (b) pure bend, in a nematic liquid crystal confined between solid plates.  A polarization is flexoelectrically induced, resulting in an electric field in the liquid crystal.  (For $e_1>0$ and $e_3>0$.)}
\end{figure}

Consider a nematic LC confined between parallel plates with surface treatment such that ${\bf n}$ lies perpendicular to the plates at the surfaces, Fig.~\ref{fig:limits}(a).
It is assumed that the LC is insulating and the thickness of the layer $l$ is sufficiently small that edge effects may be ignored.  If the plates are distorted so as to tend to form coaxial cylinders, a pure splay state of the LC is created.  In cylindrical coordinates ${\bf n}(\rho,\varphi,z)=\pm \hat{\bm\rho}$, and ${\bf S}=\hat{\bm\rho}/\rho$ ($T=0$ and ${\bf B}=0$).  For convenience, a small distortion is considered, $\rho\gg l$, such that $\rho$, and hence ${\bf S}$, are approximately constant throughout the thickness of the LC.  The mechanical energy required to distort unit volume of the LC in this way is given from Eq.~(\ref{eq:fenergy}) as
\begin{equation}
F_{\textrm{d}}=\frac{K_1^D}{2\rho^2}.
\end{equation}
(We may ignore the energy required to distort the plates, assuming it to be known and subtracted out of our considerations.)  $K_1^D$ denotes $K_1$ at constant electric displacement.\footnote{As for solid piezoelectrics (see, e.g., Ref.~\onlinecite{cady}), the electrical condition of an elastic coefficient must be specified when flexoelectricity is taken into account (see, e.g., Refs.~\onlinecite{pikin, castlesprl}).  Similarly, the mechanical conditions of a dielectric coefficient must be specified (see, e.g., Refs.~\onlinecite{cady,derzhanski,helfrich}).}

The splay creates a flexoelectric polarization ${\bf P}_\textrm{d}=e_1\hat{\bm\rho}/\rho$, which induces bound surface charge per unit area $\sigma =\pm e_1/\rho$ on the plates.  An electric field ${\bf E}=-{\bf P}_\textrm{d}/\epsilon_\parallel^a=-e_1\hat{\bm\rho}/(\epsilon_\parallel^a\rho)$ is induced in the LC, where $\epsilon_\parallel^a$ is the permittivity parallel to ${\bf n}$.  The superscript $a$ denotes $\epsilon_\parallel$ at constant curvature, following the notation of Ref.~\onlinecite{meyer}.  Energy associated with this field may do work; for example, a charged mass initially at rest within the LC will be accelerated.  Alternatively, if the plates are conducting and are subsequently electrically connected, a current will flow that may be used to drive a load.  The energy associated with unit volume of the field is
\begin{equation} \label{eq:fenergys}
F_{E}=\frac{1}{2}\epsilon_\parallel^a |{\bf E}|^2= \frac{e_1^2}{2\epsilon_\parallel^a\rho^2}.
\end{equation}
Energy conservation requires $F_{E}\leq F_{\textrm{d}}$, thus the fundamental inequalities are obtained:
\begin{subequations}
\label{inequalities}
\begin{eqnarray}
e_1^2\leq\epsilon_\parallel^a K_1^D,											\label{inequalitya} \\
\quad\textrm{and}\quad e_3^2\leq\epsilon_\perp^a K_3^D.		\label{inequalityb}
\end{eqnarray}
\end{subequations}
The second equation is derived by repeating the analysis for surface treatment such that ${\bf n}$ lies parallel to the plates, Fig.~\ref{fig:limits}(b).  A pure bend distortion is generated: ${\bf n}=\pm \hat{\bm\varphi}$, ${\bf B}=-\hat{\bm\rho}/\rho$, $F_{\textrm{d}}=K_3^D/(2\rho^2)$, ${\bf P}_\textrm{d}=-e_3\hat{\bm\rho}/\rho$, ${\bf E}=e_3\hat{\bm\rho}/(\epsilon_\perp^a\rho)$, $F_{E}=e_3^2/(2\epsilon_\perp^a\rho^2)$, where $\epsilon_\perp^a$ is the permittivity perpendicular to ${\bf n}$.  Helfrich derived inequalities very similar to (\ref{inequalities}) using alternative arguments (firstly, by considering the difference between the ``free'' and ``clamped'' dielectric coefficients, and secondly, using molecular statistical considerations).\cite{helfrich}

For LCs composed of small elongated organic molecules, typical orders of magnitude are $K_{1,3}\sim10~$pN, and $\epsilon_{\parallel,\perp}\sim 10$--$100$~pF/m.  The above inequalities~(\ref{inequalities}) then imply, to nearest order of magnitude, that $|e_{1,3}|\leq\sim 10$~pC/m.  Following the  review by Petrov,\cite{petrov} the experimental order of magnitude of $e_{1,3}$ is $10$~pC/m and the above limits are not usually violated.

Harden \textit{et al}. reported ``giant'' flexoelectricity in 2006 with a measurement of $|e_3|=62\,000$~pC/m.\cite{harden}  The material studied had dielectric and elastic coefficients comparable to conventional nematic LCs.\cite{majumdar}  Inequality~(\ref{inequalityb}) is apparently violated.  Therefore, Harden \textit{et al}.'s results are irreconcilable with the theoretical limitations established using the Oseen-Frank continuum theory of LCs and the conservation of energy.  Indeed, a number of groups have been unable to reproduce Harden \textit{et al}.'s results, using a variety of different methods, with similar,\cite{le,salter3} and identical, \cite{kumar} materials.  Molecular ``clusters'' have been invoked in recent discussion to reconcile these different results.\cite{harden,le,kumar,salter3}  Our argument is independent of such considerations; clusters are irrelevant in our conservation of energy argument.  Since $e_1$ and $e_3$ are \textit{defined within} Oseen-Frank continuum theory, it is valid to use Oseen-Frank continuum theory to assess experiments that attempt to measure them.  Harden \textit{et al}. measured an electric current produced by a mechanically distorted sample of LC confined between two flexible plates.  Our work implies that, whatever the origin of this current, it cannot constitute a measurement of $|e_3|$.  This raises the question of what \textit{did} produce it; a new interpretation would seem to be required.

More generally, important relationships for the design and development of flexoelectric materials and devices have been rigorously derived herein, in terms of Oseen-Frank continuum theory.  For liquid crystals composed of small elongated organic molecules, it seems unlikely that flexoelectric coefficients with magnitude greater than a few tens of pC/m are achievable.  According to inequalities~(\ref{inequalities}), these may only be increased if the dielectric and/or elastic coefficients are also considerably increased.  Nevertheless, molecular design protocols may produce materials for which an increased fraction of energy supplied is converted in a flexoelectric process (i.e., the limits of the inequalities may be approached more closely).  The implications of approaching this limit for distorted equilibrium structures\cite{castlesprl,porenta} suggest an interesting avenue of research.  Further, in some device configurations---such as the flexoelectric effect in chiral nematics\cite{patel,patent,broughton,coles,davidson,castlespre,chen2,carbone,salter2,castlesjsid}---the magnitude of the observed flexoelectric response is governed by a ratio of the form $|e/K|$.  Generally, the higher the ratio $|e/K|$, the lower the electric field needed to drive the device.  Inequalities~(\ref{inequalities}) suggest that, loosely, the upper limit of $|e/K|$ is increased if the dielectric coefficients are increased and/or the elastic coefficients are decreased.  Knowledge of these fundamental limits, and how they depend on material parameters, should guide and assist in the development of improved flexoelectric materials for devices.

The authors thank Daniel Corbett and Patrick Salter for useful discussions.  This work was supported by EPSRC Grants No.~EP/D04894X/1 and No.~EP/H046658/1.

\end{document}